\documentstyle[tighten,epsf,floats,preprint,aps]{revtex}
\date{January 21, 1993}
\preprint{KUNS 1180}
\newcommand\ps[2]{\epsfysize=#1\centerline{\epsffile{#2}}}
\def\lt{<}
\def\gt{>}
\catcode`<=13
\catcode`>=13
\newcommand<{\langle}
\newcommand>{\rangle}
\newcommand\carbon{{{}^{12}{\rm C}}}
\newcommand\Be[1]{{}^{#1}{\rm Be}}
\renewcommand\vec[1]{{\bf #1}}
\begin{document}
\title{
  Momentum distribution of fragments in heavy ion reactions:\\
  Dependence on the stochastic collision process
}
\author{Akira Ono, Hisashi Horiuchi and Toshiki Maruyama}
\address{
  Department of Physics, Kyoto University, Kyoto 606-01, Japan
}
\author{Akira Ohnishi}
\address{
  Research Center for Nuclear Physics, Osaka University,
  Ibaraki 567, Japan
}
\maketitle
\begin{abstract}
Momentum distribution of fragments produced in $\carbon+\carbon$ 
reaction at 28.7 MeV/nucleon is analyzed with antisymmetrized 
version of molecular dynamics (AMD).
Calculations are made for several cases of stochastic collisions
and the projectile fragmentation peak in momentum distribution is 
reproduced by the incorporation of many-body nature in stochastic 
collision process.
Furthermore the value of the cross section is found to be reflected 
in the low momentum part of fragments such as alpha particles and 
$\Be9$.
\end{abstract}
\narrowtext

\section{Introduction}
Fragment formation is the most characteristic feature of the heavy 
ion reactions in the intermediate energy region.
The main reaction mechanism of the peripheral collision is the 
projectile fragmentation, in which a projectile-like fragment 
produced by stripping some nucleons from the projectile is emitted 
with the velocity close to the beam velocity.
In the central collision, on the other hand, the reaction mechanism 
is more complicated and many fragments with various mass numbers 
are produced.
At the same time, if we focus on the collective momentum flow 
instead of fragment mass number, the central collision in this 
energy region is believed to be a good tool to investigate the 
equation of state of nuclear matter.

There are many models which describe these features of intermediate 
energy heavy ion reactions.
Abrasion-ablation model, for example, gives a very intuitive 
picture of projectile fragmentation and is successful in 
reproducing the mass distribution of projectile-like fragments.
This model, however, assumes the reaction mechanism itself, and 
cannot answer the question whether the projectile fragmentation is 
possible in low energy heavy ion reactions.
In these years, projectile fragmentation is observed in experiments 
at the incident energies down to 30 MeV/nucleon.
It may not be trivial theoretically that participant spectator 
picture is true at this incident energy comparable to the Fermi 
energy of nuclei.
The central collisions, on the other hand, are sometimes described 
by the statistical models if one put emphasis on the 
multifragmentation as the phase transition.
In the statistical models, however, dynamical effects are not fully 
incorporated and collective momentum flow is usually calculated by 
VUU approach which cannot describe fragment formation without 
further assumptions.

In order to study these phenomena in a unified way with few 
assumptions on reaction mechanisms, microscopic simulations which 
can treat fragment formation is indispensable.
Quantum molecular dynamics (QMD) \cite{AICH,MARUa} has been the 
most powerful among such microscopic simulation methods.
QMD is, however, largely of classical nature and, for example, it 
cannot describe shell effects neither of colliding individual 
nuclei nor in the reaction process.
In the previous work \cite{ONOa,ONOb}, we constructed an 
antisymmetrized version of molecular dynamics (AMD) by 
incorporating two-nucleon collision process as the residual 
interaction into the fermionic molecular dynamics (FMD) \cite{FELD} 
proposed by Feldmeier.
AMD describes the system with a Slater determinant of Gaussian wave 
packets, and therefore can describe quantum mechanical features 
such as shell effects.
Furthermore AMD can describe fragment formation as well as QMD.
These features of AMD has been demonstrated by the calculation of 
mass distribution of fragments in the previous work 
\cite{ONOa,ONOb} with the combination to the statistical cascade 
decay calculation.
AMD can also be applied to the study of nuclear structure of light 
nuclei \cite{HORIa,KANADAa} with some extensions of the wave 
function.

In this paper, the momentum distribution of fragments produced in 
$\carbon+\carbon$ reaction at 28.7 MeV/nucleon is mainly discussed 
because it has much more information on the reaction mechanisms 
such as projectile fragmentation than the mass distribution does.
Since the incorporation of the stochastic collision process has 
some ambiguity,
we have made calculations with different types of stochastic 
collision process,
not only by changing the cross section but also by including 
many-body nature such as nucleon-alpha collisions in addition to 
the usual two-nucleon collisions.
It will turn out that the momentum distribution of fragments is 
very sensitive to the stochastic collision process.
This fact means that detailed analysis of momentum distribution of 
fragments is powerful to settle the ambiguity of stochastic 
collision process which has been the main obstacle in extracting 
the equation of state of nuclear matter by the analysis of 
collective momentum flow.

In Sec.\ II, the framework of AMD, with the inclusion nucleon-alpha 
collisions in addition to the usual two-nucleon collisions, is 
presented.
In Sec.\ III, the results of calculation for the reaction 
$\carbon+\carbon$ are compared to the data, and the dependence of 
the momentum distribution of fragments on the stochastic collision 
process is discussed, which will reveal the reaction mechanisms 
both of the peripheral collision and of the central collision.
The manifestation of collective momentum flow of fragments will 
also be mentioned.
Finally in Sec.\ IV, we will summarize the results and discuss 
future problems.

\section{Formulation of AMD}
Since the framework of the antisymmetrized version of molecular 
dynamics (AMD) was described in detail in Ref.\ \cite{ONOb},
here is shown only the outline of our framework about the wave 
function and the equation of motion.
The treatment of the stochastic collision process is updated from 
Ref.\ \cite{ONOb}, and therefore is described here in detail.

\subsection{Wave function}

In AMD, the wave function of $A$-nucleon system $|\Phi>$ is
described by a Slater determinant $|\Phi(\vec Z)>$;
\begin{equation}
  |\Phi(\vec Z)>={1\over\sqrt{A!}}\det\Bigl[\varphi_i(j)\Bigr],\quad
   \varphi_i=\phi_{\vec Z_i}\chi_{\alpha_i},
\end{equation}
where $\alpha_i$ represents the spin-isospin label of $i$-th single
particle state, $\alpha_i={\rm p}\uparrow$, ${\rm p}\downarrow$,
${\rm n}\uparrow$, or ${\rm n}\downarrow$, and $\chi$ is the 
spin-isospin
 wave function. $\phi_{\vec Z_i}$ is the spatial wave
function of $i$-th single particle state which is a Gaussian wave 
packet
\begin{equation}
  <\vec r\,|\phi_{\vec Z_i}>=\Bigl({2\nu\over\pi}\Bigr)^{3/4}
  \exp\biggl[-\nu\Bigl(\vec r-{{\vec Z}_{i}\over\sqrt\nu}\Bigr)^2
                      +{\textstyle{1\over2}}{\vec Z}_{i}^2\biggr],
\end{equation}
where the width parameter $\nu$ is treated as time-independent in 
our model.
We took $\nu=0.16$ ${\rm fm}^{-2}$ in the calculation presented in 
this paper.
If we define $\vec D$ and $\vec K$ as
$\vec Z = \sqrt\nu\vec D + (i/2\hbar\sqrt\nu)\vec K$,
then
\begin{eqnarray}
<\phi_{\vec Z}|\vec r\,|\phi_{\vec Z}>
/<\phi_{\vec Z}|\phi_{\vec Z}>&=&\vec D,\\
<\phi_{\vec Z}|\vec p\,|\phi_{\vec Z}>
/<\phi_{\vec Z}|\phi_{\vec Z}>&=&\vec K.\\
\nonumber
\end{eqnarray}

\subsection{Equation of motion}

The time developments of the centers of
Gaussians, $\{\vec Z_i (i=1,2,\ldots,A)\}$,
are determined by two processes.
One is the time development determined by the time-dependent 
variational principle
\begin{equation}
  \delta\int_{t_1}^{t_2}dt\,
  { <\Phi(\vec Z)|(i\hbar{d\over dt}-H)|\Phi(\vec Z)>
   \over<\Phi(\vec Z)|\Phi(\vec Z)>}=0,
\end{equation}
which leads to the equation of motion for $\{\vec Z\}$;
\begin{equation}
  i\hbar\sum_{j\tau}C_{i\sigma,j\tau}\dot Z_{j\tau}=
  {{\partial{\cal H}}\over{\partial Z_{i\sigma}^*}}
  \qquad \hbox{and c.c.},
\end{equation}
where $\sigma, \tau=x,y,z$.
${\cal H}$ is the expectation value $<H>$ of quantum mechanical 
Hamiltonian $H$,
\begin{equation}
  {\cal H}(\vec Z,\vec Z^*)={<\Phi(\vec Z)|H|\Phi(\vec Z)>
                              \over<\Phi(\vec Z)|\Phi(\vec Z)>} ,
\end{equation}
and
\begin{equation}
  C_{i\sigma,j\tau}
  =  {\partial^2\over\partial Z_{i\sigma}^*\partial Z_{j\tau}}
      \log<\Phi(\vec Z)|\Phi(\vec Z)>
\end{equation}
is a positive definite hermitian matrix.

The same effective interaction as in the previous work 
\cite{ONOa,ONOb}, i.e., Volkov No.1 \cite{VOLKOV} with $m=0.576$ is 
used.
Coulomb interaction is included.
The problem of the zero-point oscillation of the fragment 
center-of-mass motion is overcome phenomenologically with the 
prescription we have proposed in Ref.\ \cite{ONOb}.
As has been shown in Ref.\ \cite{ONOb}, the binding energies of 
nuclei lighter than $\carbon$ are reproduced very well by this 
choice of parameters concerned with equation of motion.

\subsection{Stochastic collision process}

The second process which determines the time development of the 
system is the stochastic collision process due to the residual 
interaction.
We incorporate this process in the similar way to QMD, i.e., the 
momenta of two nucleons which have approached toward each other are 
changed stochastically.
In AMD, however, this is not straightforward to do since the 
centers of Gaussian wave packets $\{\vec Z_i\}$ are no more 
physical coordinates of nucleons due to the effect of 
antisymmetrization.
We overcome this difficulty by introducing the physical coordinates 
$\{\vec W_i\}$ \cite{ONOa,ONOb} as
\begin{equation}
  \vec W_i=\sum_{j=1}^A \Bigl(\sqrt Q\Bigr)_{ij}\vec Z_j,
\end{equation}
where
\begin{equation}
  Q_{ij}={\partial\over\partial(\vec Z_i^*\cdot\vec Z_j)}
         \log <\Phi(\vec Z)|\Phi(\vec Z)>.
\end{equation}
The position ${\vec R}_j$ and momentum ${\vec P}_j$ of
${\vec W}_j = \sqrt\nu{\vec R}_j + (i/2\hbar\sqrt\nu){\vec P}_j$
can be interpreted to be physical position and momentum coordinates 
of nucleons \cite{ONOb}.

As a manifestation of Pauli principle, there is Pauli-forbidden 
region in the phase space of physical coordinates.
Taking account of this fact, the Pauli-blocking in the final state 
of two-nucleon collision is introduced in a natural way 
\cite{ONOa,ONOb}, i.e., the collision is Pauli-blocked if the 
physical coordinate of the final state is in the Pauli-forbidden 
region.

The in-medium cross section of two-nucleon collision is 
parametrized as
\begin{equation}
\sigma_{NN}
={100\,{\rm mb}\over1+E/200\,{\rm MeV}+C\min((\rho/\rho_0)^2,1)},
\label{DefOfsigma}
\end{equation}
where $E$ is the energy in the laboratory system of two-nucleon 
scattering, and $\rho$ is the density at the middle point of 
colliding two nucleons.
The parameter $C$ controls the reduction of the cross section due 
to the medium effect.
Isotropic scattering is assumed.

So far we considered only two-nucleon collisions as the effect of 
residual interaction.
There may be, however, other effects of residual interaction in 
which more than two nucleons are concerned.
A simple example which shows the necessity of the many-body nature 
of residual interaction in the AMD framework is the proton-alpha 
elastic scattering which is the dominant process in $p+\alpha$ 
reaction up to the intermediate energy region.
First of all, the incident proton is deflected at most $20^\circ$ 
by the equation of motion of AMD if the incident energy is greater 
than 30 MeV, though the large angle elastic scattering is observed 
in experiments.
The two-nucleon collision process in AMD does not remedy this 
situation
because the alpha particle is inevitably excited once the 
two-nucleon collision occurs between the incident proton and a 
nucleon in the alpha particle.

\begin{figure}
\ps{9cm}{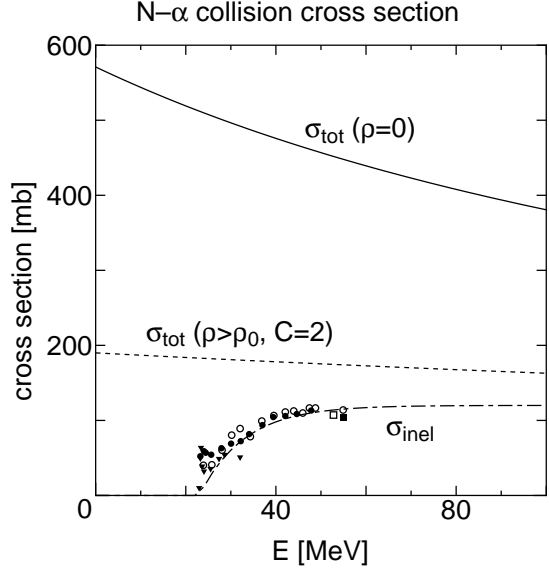}
\caption{\label{SIGNALPHA}
Energy dependence of the cross sections of nucleon-alpha collisions.
Solid line and dotted line are the total cross sections 
parametrized as Eq.\ (\protect\ref{DefOfSigmaTot}), for $\rho=0$ 
and $\rho\ge\rho_0$ ($C=2$) respectively.
Dot dashed line is the inelastic cross section parametrized as Eq.\ 
(\protect\ref{DefOfSigmaInel}) and various symbols are the 
experimental data of reaction cross section of $p+\alpha$ reaction 
presented in Ref.\ \protect\cite{HOUDAYER}.
}
\end{figure}

\begin{figure}
\ps{9cm}{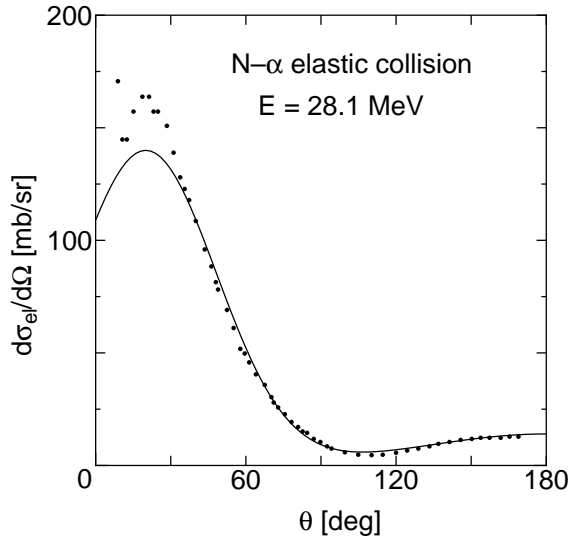}
\caption{\label{DSDO}
Differential cross section of nucleon-alpha elastic collisions at 
$E=28.1$ MeV.
Solid line shows the parametrization of Eq.\ 
(\protect\ref{DefOfDSDO}) and dots are the data 
\protect\cite{HOUDAYER}.
}
\end{figure}

In order to study the effect of many-body nature of residual 
interaction in heavy ion reactions, we also incorporate following 
many-body stochastic collisions, which we call nucleon-alpha 
collisions, in addition to the usual two-nucleon collisions.
When a nucleon and an alpha cluster (a cluster of four physical 
coordinates with different spin-isospins, which is judged by the 
chain clustering method with the critical distance $|\Delta\vec 
W|=0.25$) have approached toward each other,
(i) they are scattered elastically with the cross section 
$\sigma_{N\alpha,\rm el}(E,\rho)$ and (ii) the nucleon and a 
nucleon in the alpha cluster are scattered in the usual way with 
the cross section $\sigma_{N\alpha,\rm inel}(E)$.
The same energy and density dependence as that of the two-nucleon 
collision is assumed for the in-medium total nucleon-alpha cross 
section as
\begin{eqnarray}
\sigma_{N\alpha,\rm tot}(E,\rho)&=&
\sigma_{N\alpha,\rm el}(E,\rho)
+\sigma_{N\alpha,\rm inel}(E)\nonumber\\
&=&{571\,{\rm mb}\over1+E/200\,{\rm MeV}
+C\min((\rho/\rho_0)^2,1)}\nonumber\\     \label{DefOfSigmaTot}
\end{eqnarray}
As for the energy dependent inelastic cross section, we use the 
experimental data of reaction cross section \cite{HOUDAYER} (see 
Fig.\ \ref{SIGNALPHA}) which can be parametrized as
\begin{equation}
\sigma_{N\alpha,\rm inel}(E)=
  \max\Bigl(
    120-162e^{-(E-20\,{\rm MeV})/10\,{\rm MeV}},\,
    0
  \Bigr)\,{\rm mb}.                      \label{DefOfSigmaInel}
\end{equation}
The angular distribution of the elastic nucleon-alpha scattering is 
assumed to be
\begin{equation}
{d\sigma_{N\alpha,\rm el}\over d\Omega}\propto
\exp\biggl[-\Bigl({180^\circ-\theta\over70^\circ}\Bigr)^2\biggl]
+10\exp\biggl[-\Bigl({\theta-20^\circ\over40^\circ}\Bigr)^2\biggl].
                                         \label{DefOfDSDO}
\end{equation}
The parametrization of Eqs.\ (\ref{DefOfSigmaTot}), 
(\ref{DefOfSigmaInel}) and (\ref{DefOfDSDO}) reproduces the data of 
elastic differential cross section of $p+\alpha$ reaction at 
$E=28.1$ MeV \cite{HOUDAYER} for $\theta \ge 30^\circ$ in the case 
$\rho=0$ as shown in Fig.\ \ref{DSDO}.
Note that it is not necessary to reproduce the data for small 
scattering angles since the small angle elastic scatterings are 
described not only by the stochastic collisions but also by the 
equation of motion of AMD.
The energy dependences of total and inelastic cross sections are 
shown in Fig.\ \ref{SIGNALPHA}, where the data of the reaction 
cross section are also shown.
The parametrization is consistent with the essential feature that 
the elastic cross section decreases as the energy increases.

When the alpha cluster composed of nucleons $i_1$, $i_2$, $i_3$ and 
$i_4$ is to be scattered elastically, the momentum transfer which 
should be received by the alpha cluster is shared not only by the 
four nucleons but also by the other nucleons $j$ with the weights 
proportional to $\exp(-4|\vec Z_j-\vec Z_{i_\alpha}|^2)$, where the 
nucleons $j$ and $i_\alpha$ have the same spin and isospin.
This prescription is aimed to suppress the Pauli-blocking 
probability of the scattered alpha cluster which seems too large 
but for this prescription.
When two alpha clusters have approached, nucleon-alpha collisions 
defined above are made between nucleons in one alpha cluster and 
the other alpha cluster.
It is randomly decided which alpha cluster is treated as merely 
four nucleons.

Although a procedure to scatter two particles with the given cross 
section $\sigma$ has been described in Ref.\ \cite{ONOb}, the 
scattering probability appearing in this procedure exceeds unity 
when it is applied to the nucleon-alpha collisions.
In the calculation presented in this paper, we use the following 
procedure when nucleon-alpha collision process is included.
At each time step, it is examined whether each pair of two 
particles (which may be two nucleons or a nucleon and an alpha 
cluster) should make a stochastic collision.
The probability with which the two particles are scattered (up to 
the Pauli-blocking) when the relative coordinate changes from $\vec 
r$ to $\vec r+d\vec r$ in a time interval $dt$ is assumed to be 
proportional to the density overlap of the two Gaussian wave 
packets as
\begin{equation}
P(\vec r)|d\vec r| = \alpha e^{-\nu\vec r^2}|d\vec r|,
\end{equation}
where the proportionality constant $\alpha$ is determined by the 
condition that the total cross section should be $\sigma$ if two 
particles move on straight lines before they are scattered,
\begin{equation}
2\pi\int bdb
\Bigl[1-\exp\Bigl(-\alpha\sqrt{\pi/\nu}e^{-\nu b^2}\Bigr)\Bigr]
= \sigma.
\end{equation}
This condition can be rewritten as
\begin{equation}
y=-\sum_{n=1}^\infty{(-x)^n\over n\cdot n!},
\end{equation}
where
\begin{equation}
y={\nu\sigma\over\pi},\quad
x=\alpha\sqrt{\pi\over\nu}.
\end{equation}
When more than two stochastic collisions should occur at the same 
time step, the order of collisions is decided randomly.
\section{Results and discussions}
\subsection{Numerical calculation}

AMD simulations are made for the reaction $\carbon+\carbon$ at the 
incident energy 28.7 MeV/nucleon.
The initial state of each simulation is constructed by boosting 
randomly rotated two $\carbon$ nuclei in ground states which have 
been obtained by frictional cooling method \cite{HORIa,ONOa,ONOb}.
Simulations are repeated for many times (typically 1000 times) for 
various impact parameters and for different random seeds, and each 
simulation is considered to correspond to an experimental event.

AMD calculations are made for three types of stochastic collision 
process.
The calculation presented in Refs.\ \cite{ONOa,ONOb}, where only 
the two-nucleon collisions are incorporated and the parameter $C$ 
in Eq.\ (\ref{DefOfsigma}) which controls the medium effect is 
taken to be 0, is called the case (a) here.
In the cases (b) and (c), both the two-nucleon collisions and 
nucleon-alpha collisions are incorporated, and $C=1$ for the case 
(b), and $C=2$ for the case (c).
Note that the cross section is smaller for larger value of $C$.

After the AMD calculation which is truncated at $t\approx200$ 
fm/$c$, most of the produced fragments are in their excited states 
and should decay into lighter fragments before they are detected in 
experiment.
We calculate statistical cascade decay of each of these fragments 
by using a code \cite{MARUb,ONOa,ONOb} which is similar to CASCADE 
of P\"uhlhofer \cite{PUHLHOFER}.

\subsection{Mass distribution}
\begin{figure}
\ps{10cm}{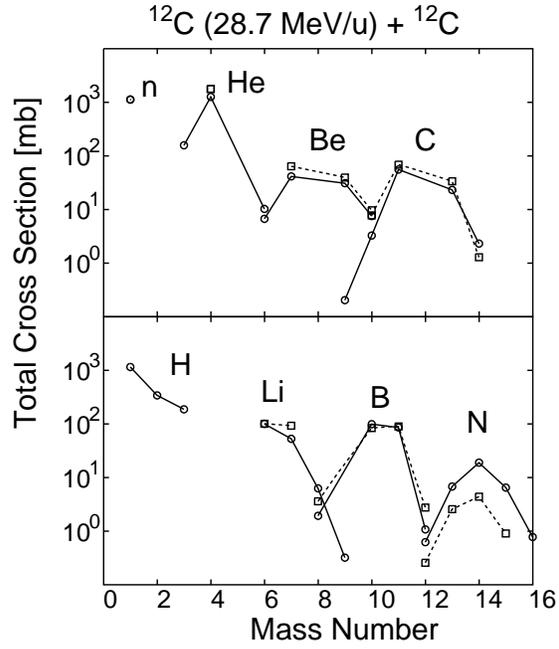}
\caption{\label{IDST}
Isotope distribution of produced fragments.
The results of calculation in the case (c) are shown by circles 
connected by solid lines, and the data are shown by squares 
connected by dotted lines.
}
\end{figure}
\begin{figure}
\vspace{-2cm}
\ps{10cm}{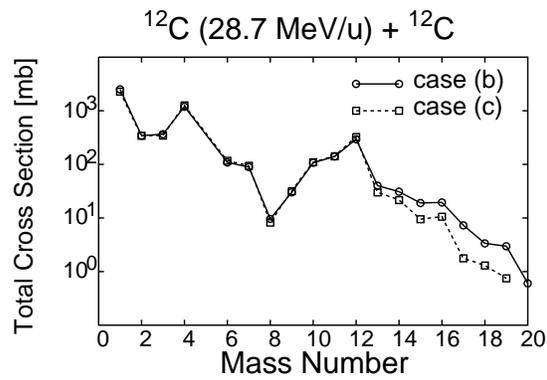}
\vspace{-1cm}
\caption{\label{MDST}
Calculated mass distribution for the cases (b) and (c) of 
stochastic collision process.
}
\end{figure}
The calculated isotope distribution of fragments is compared to the 
experimental data in Fig.\ \ref{IDST} for the case (c).
As in the case (a) which we have already discussed in Refs.\ 
\cite{ONOa,ONOb}, the mass distribution is reproduced very well, 
especially the large production cross section of alpha particles.
{}From Fig.\ \ref{MDST} in which mass distributions for two 
stochastic collision processes (b) and (c) are compared, we can see 
that the mass distribution of fragments lighter than the projectile 
and the target is insensitive to the stochastic collision process.
The cross sections of heavier fragments are, on the other hand, 
reflect the stochastic collision cross section.
The smaller the parameter $C$ is, i.e., the larger the cross 
section of stochastic collisions is, the more abundantly heavy 
fragments are produced.

\subsection{Projectile fragmentation}

\begin{figure}
\ps{10cm}{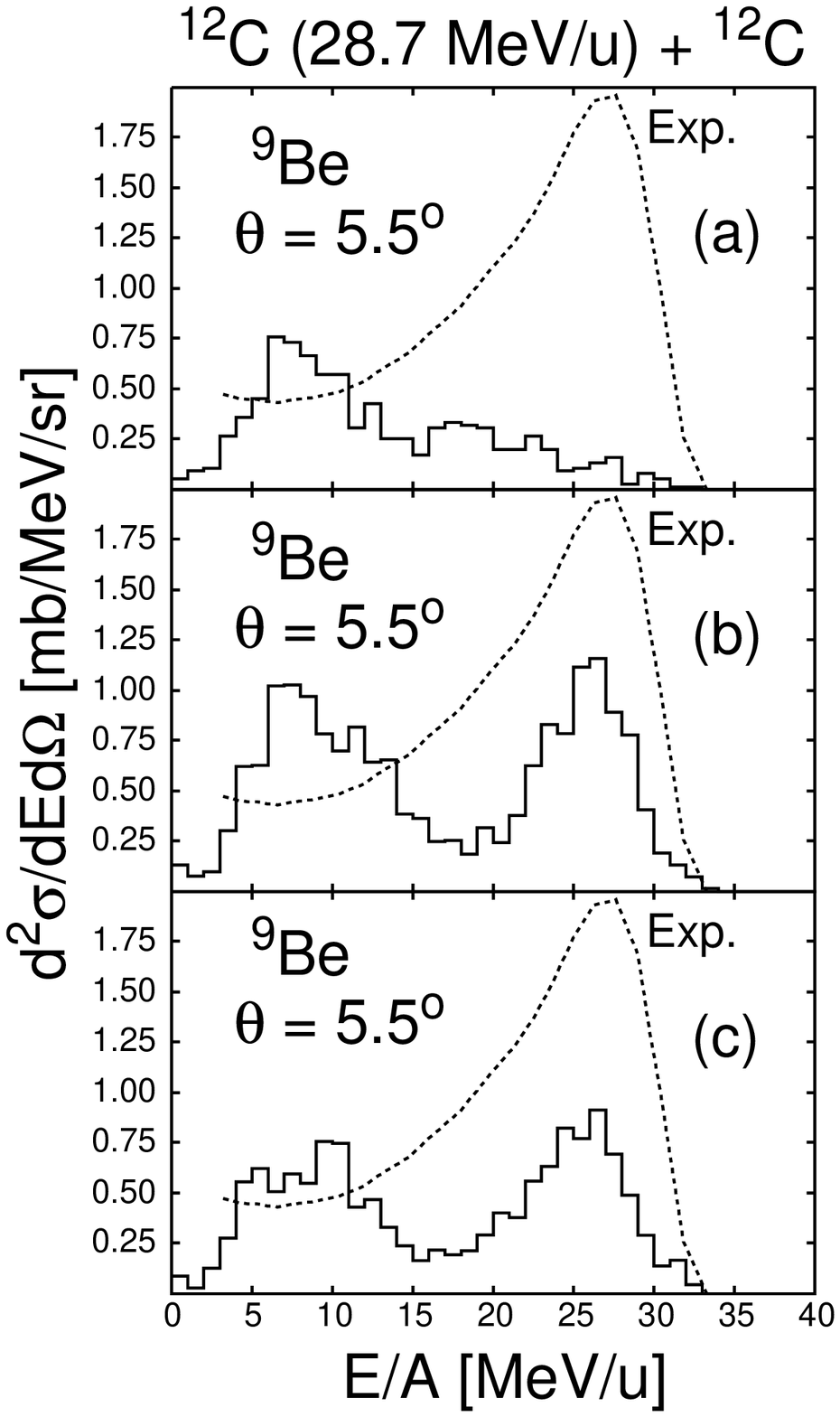}
\caption{\label{EBe}
Energy spectra of $\Be9$ at $\theta=5.5^\circ$ for three cases (a), 
(b) and (c) of stochastic collision process.
Histograms are the result of calculation and the dotted curves are 
the data.
}
\end{figure}
In Fig.\ \ref{EBe}, the energy spectra of $\Be9$ at the angle 
$5.5^\circ$ are compared to the experimental data \cite{CZUDEK} for 
three cases of stochastic collisions.
In experiment, the projectile fragmentation peak appears very close 
to the beam velocity even in this relatively low incident energy 
$E=28.7$ MeV/nucleon, but the calculation (a) only with two-nucleon 
collisions fails to explain this phenomenon.
This failure can be understood by the fact that the momentum of the 
nucleon in the projectile which has experienced a two-nucleon 
collision with a nucleon in the target is too small in the 
projectile frame for the nucleon to escape from the projectile.
If two nucleons which make a collision have zero momenta in the 
projectile and the target frame respectively and the scattering 
angle is $90^\circ$, which is the most probable case, the nucleon 
in the projectile has the kinetic energy $E/2$ in the projectile 
frame (as long as the energy correction \cite{ONOb} in the 
two-nucleon collision is ignored), where $E$ is the incident energy 
per nucleon in the laboratory system.
Since $E=28.7$ MeV/nucleon in the present case, the typical kinetic 
energy of scattered nucleon in the projectile is 15 MeV.
It is difficult for this scattered nucleon to escape from the 
projectile since the proton separation energy of $\carbon$ is also 
about 15 MeV.

This situation evidently depends on the incident energy.
In fact, in the calculation with the incident energy 70 MeV/nucleon 
and the impact parameter 6 fm, and only with two-nucleon 
collisions, we have found that 15 projectile-like and target-like 
fragments with mass number 11, 10 and 9 have been produced in 58 
events before statistical cascade decay, while only one 
projectile-like fragment is found in the same calculation but with 
the incident energy 28.7 MeV/nucleon.

\begin{figure}
\vspace{-2cm}
\ps{10cm}{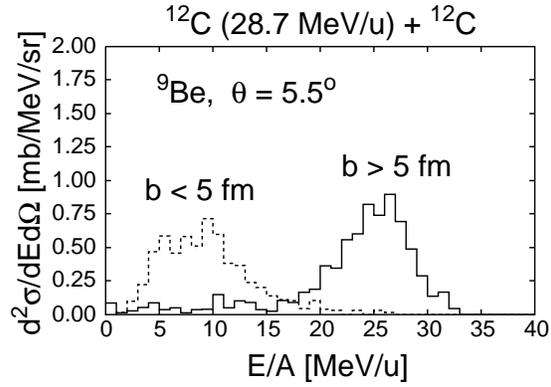}
\vspace{-1cm}
\caption{\label{EBeBIMP}
Calculated contribution to the energy spectrum of $\Be9$ at 
$\theta=5.5^\circ$ from peripheral ($b\gt5$ fm) and central 
($b\le5$ fm) collisions in the case (c) of stochastic collision 
process.
}
\end{figure}
\begin{figure}
\ps{10cm}{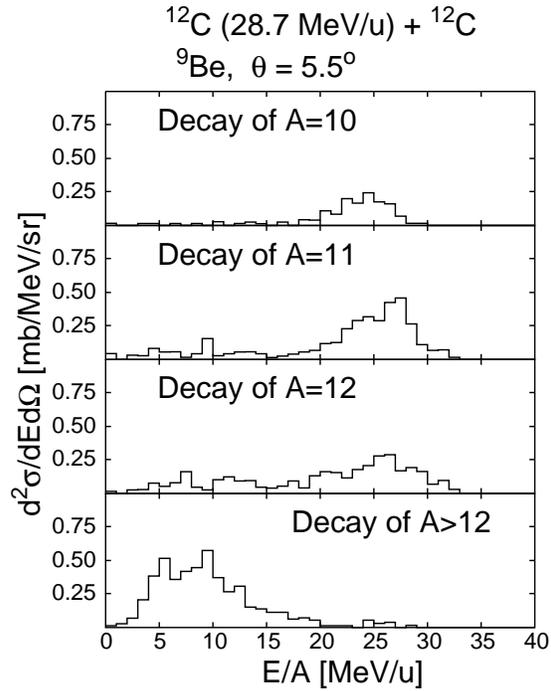}
\caption{\label{EBeDECAY}
Contribution to the energy spectrum of $\Be9$ at $\theta=5.5^\circ$ 
from statistical cascade decay of excited fragments, with mass 
numbers 10, 11, 12 and more, produced in AMD calculation for the 
case (c) of stochastic collision process.
}
\end{figure}
The inclusion of many-body nature in stochastic collision process 
has the effect to increase the kinetic energy of scattered nucleon 
in the projectile, and therefore the nucleon can easily escape from 
the projectile or, otherwise, the excitation energy of the 
projectile is large enough for the projectile to decay into $\Be9$.
By the inclusion of nucleon-alpha collisions, we have succeeded in 
the reproduction of projectile fragmentation peak in energy 
spectrum (Fig.\ \ref{EBe}), at least qualitatively.
We have checked that the calculated projectile fragmentation peak 
comes from the impact parameter range $b\gt5$ fm as shown in Fig.\ 
\ref{EBeBIMP}, and in Fig.\ \ref{EBeDECAY} the effect of 
statistical cascade decay on the energy spectrum of $\Be9$ is shown.
It is seen that most of the projectile-like $\Be9$ fragments are 
not produced directly in AMD calculation but they are the decay 
products of excited fragments with mass numbers 10, 11 and 12.

\subsection{Low energy part of spectrum}
\begin{figure}
\ps{10cm}{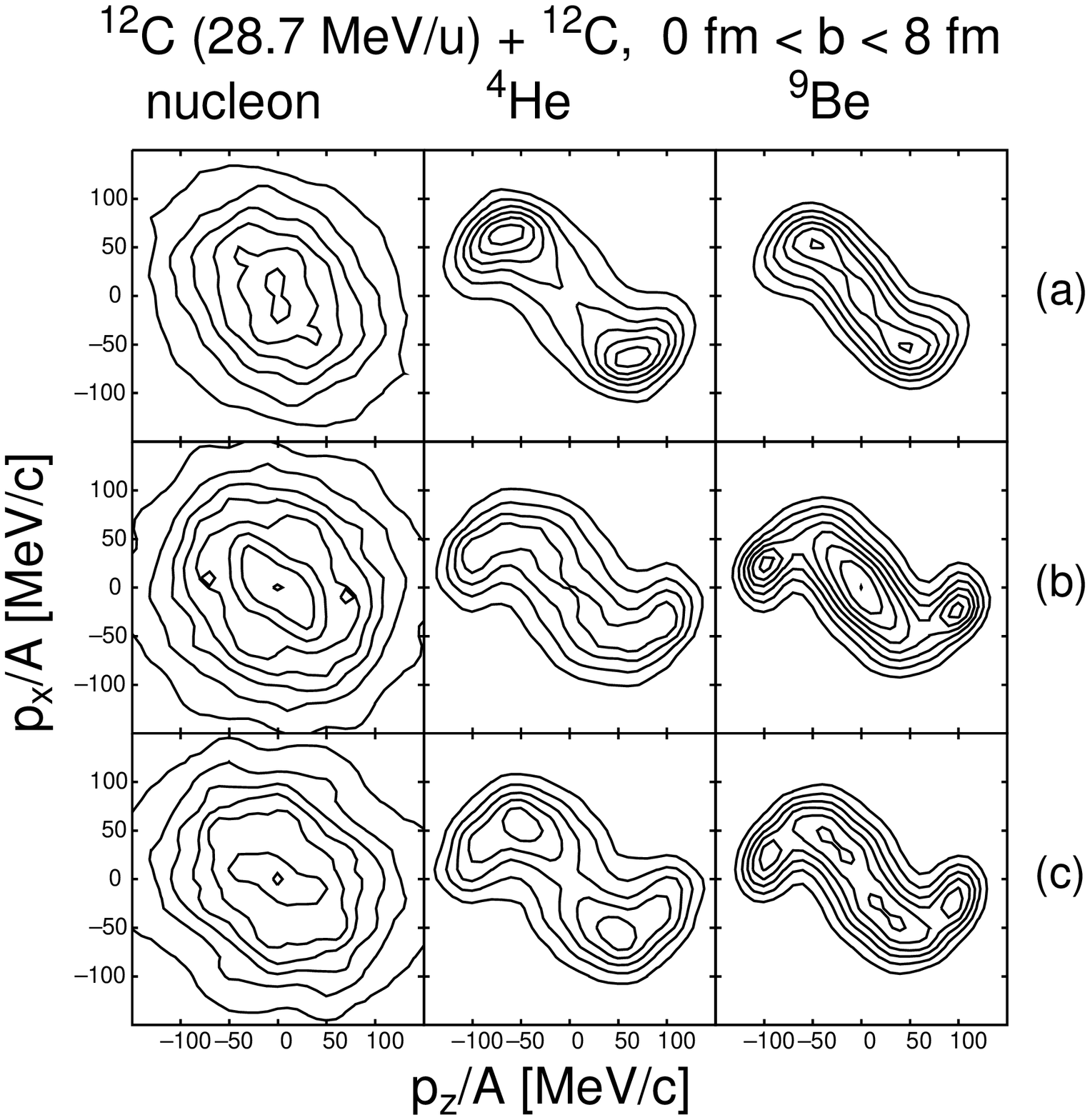}
\caption{\label{MOMCNT}
Contour maps of distribution of momentum component in the reaction 
plane of nucleons, alpha particles and $\Be9$ fragments for three 
cases (a), (b) and (c) of stochastic collision process.
The $z$-axis is the beam direction and the $zx$-plane is the 
reaction plane.
The distribution has been smoothed out by the Gaussian distribution 
with the standard deviation $\Delta p/A=10$ MeV/$c$.
}
\end{figure}
\begin{figure}
\ps{10cm}{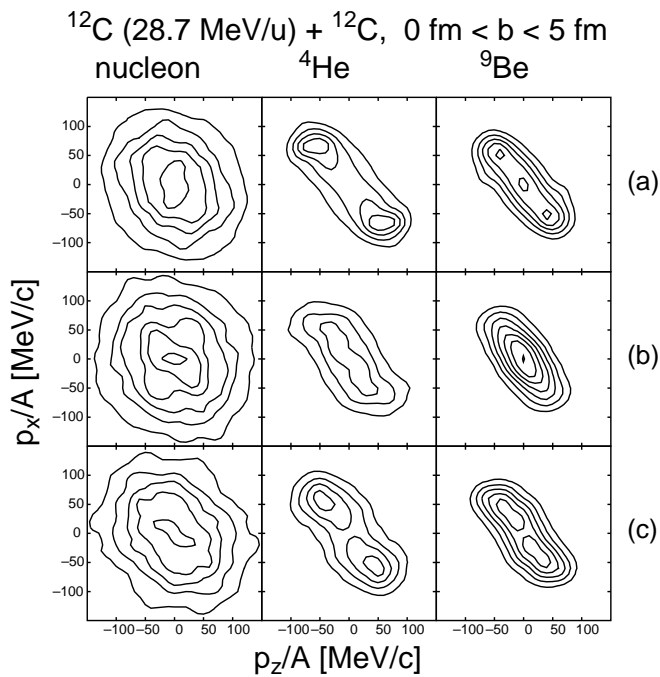}
\caption{\label{MOMCNTc}
The same as Fig.\ \protect\ref{MOMCNT}, but with the contribution 
only from central collisions ($b\lt5$ fm).
}
\end{figure}
In the energy spectrum of $\Be9$ (Fig.\ \ref{EBe}), there has 
appeared another bump in the low energy part in our calculation.
This bump has turned out to come from central events with impact 
parameters smaller than 5 fm (see Fig.\ \ref{EBeBIMP}) and 
therefore has been produced by completely different reaction 
mechanism from the projectile fragmentation.
{}From Fig.\ \ref{EBeDECAY}, it is seen that this bump is due to 
the decay of compound nuclei with mass number larger than 12.
The height of this bump is sensitive to the cross section of 
stochastic collisions as can be seen from Fig.\ \ref{EBe}.
In Fig.\ \ref{MOMCNT}, calculated distributions of the momentum 
component in the reaction plane of various fragments are shown as 
contour maps for three types of stochastic collision process.
We can see the projectile fragmentation peak in the cases (b) and 
(c).
In addition to this peak, there has appeared another hill which 
comes from central events as can be seen by the comparison with 
Fig.\ \ref{MOMCNTc} which shows the contribution from impact 
parameter range $b\lt5$ fm.
The shape of this hill is dependent on the cross section of 
stochastic collisions.
It is concentrated at the origin in the case (b), while it is 
separated into two mountains in the cases (c) and (a).
This feature is, of course, responsible for the cross section 
dependence of the low energy part in Fig.\ \ref{EBe}, and is 
understandable by the idea that the dissipation of incident kinetic 
energy of projectile and target in central events is smaller for 
smaller cross section of stochastic collisions.
As long as the central events are concerned, the cases (a) and (c) 
produce similar results.

We can see similar dependence on stochastic collision process of 
the shape of the hill in the momentum distribution of alpha 
particles in Fig.\ \ref{MOMCNT} and Fig.\ \ref{MOMCNTc}.
This means that the momentum distribution of fragments such as 
alpha and $\Be9$ is a good indicator of the cross section of 
stochastic collisions in the reaction $\carbon+\carbon$,
though this situation may depend on the mass number of target since 
the dissipation of kinetic energy of projectile is expected to be 
large for heavy target.
The momentum distribution of nucleons, on the other hand, does not 
show clear dependence on the cross section, while the spreading of 
distribution is larger if nucleon-alpha collisions are included.
The last feature is considered to appear because the momentum of 
nucleon which has been scattered by an alpha cluster is larger than 
the momentum of nucleon which has been scattered by another nucleon.
It is desirable to have data of nucleon energy spectra in this 
reaction.

\subsection{Discussion on the collective momentum flow}

\begin{figure}
\ps{9cm}{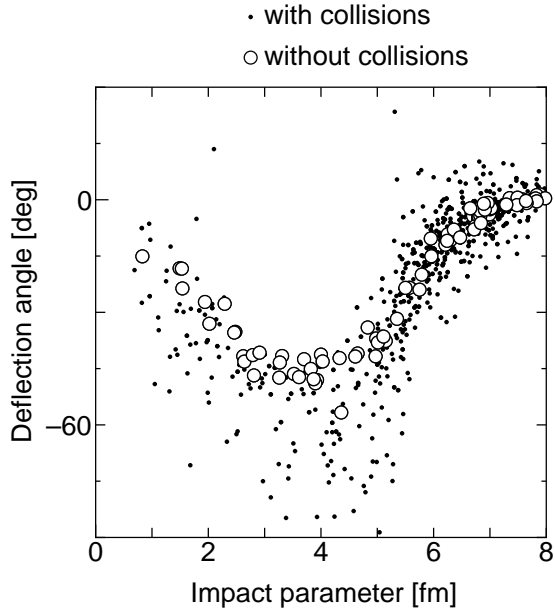}
\caption{\label{DEFANG}
Deflection angle of the $\carbon$ fragments produced in AMD 
calculation before statistical cascade decay as a function of 
impact parameter.
Open circles show the result of calculation without any stochastic 
collisions, and dots show the result of the calculation for the 
case (c) of stochastic collision process.
}
\end{figure}
\begin{figure}
\vspace{-2cm}
\ps{10cm}{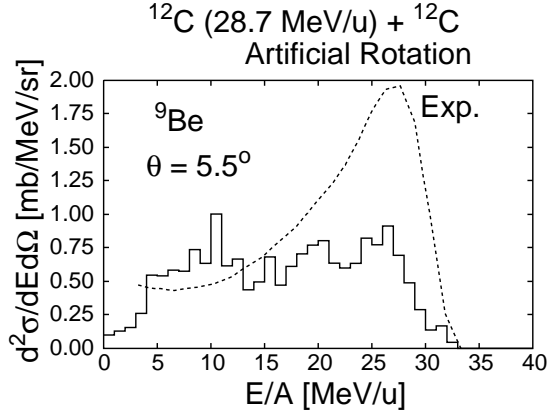}
\vspace{-1cm}
\caption{\label{EBeROT}
Energy spectrum of $\Be9$ at $\theta=5.5^\circ$ obtained by 
artificially rotating the momenta by $+20^\circ$ in the reaction 
plane when $|p_z/A|\lt75$ MeV/$c$.
}
\end{figure}

In the calculated energy spectrum of $\Be9$ at the forward angle 
(Fig.\ \ref{EBe}), there has appeared a dip, which does not exist 
in the data, between the projectile fragmentation peak and the low 
energy bump.
This dip corresponds to the valley in the momentum distribution 
(Fig.\ \ref{MOMCNT}) which has appeared as the result of clear 
separation of two components, i.e., projectile fragmentation peak 
and the low momentum hill.
The direction of the ridge of the hill in the momentum 
distributions of fragments is about $-45^\circ$, which means that 
the projectile and the target go around each other and are 
deflected by this angle on an average.
This clear flow pattern in calculation can be understood from Fig.\ 
\ref{DEFANG}, where the emitted angles of $\carbon$ fragments 
before statistical decay are plotted as a function of impact 
parameter for two calculations with and without collisions.
We can see that the flow angle, about $-45^\circ$, has been caused 
not by the effect of stochastic collisions but by the mean field 
effect, and the deflection angle suddenly changes from $-10^\circ$ 
to $-45^\circ$ when the impact parameter changes from 6 fm to 5 fm.
If this change were more moderate and/or the negative flow were 
smaller, the separation between two component in the momentum 
distribution would not have been so clear and the valley in the 
energy spectrum of $\Be9$ would have been filled up.
In fact, we have checked that if the Gogny force \cite{GOGNY} is 
used instead of the Volkov force, the calculation without 
collisions gives the negative flow angle which is about half of 
that with the Volkov force.
We get the energy spectrum of $\Be9$ shown in Fig.\ \ref{EBeROT} by 
artificially rotating the momenta of produced fragments by 
$+20^\circ$ in the reaction plane when $|p_z/A|\lt75$ MeV/$c$.

\begin{figure}
\ps{10cm}{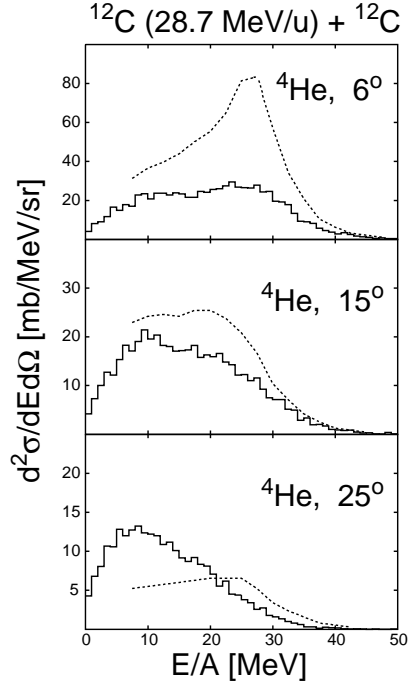}
\caption{\label{EHe}
Energy spectra of alpha particles.
Results of calculation are shown by histograms, and the data are 
shown by dotted curves.
Alpha particles produced in the calculation are assumed to have 
momentum spreading of Gaussian form with the standard deviation 
$\Delta p/2$, where $\Delta p$ is the momentum spreading of the 
wave packet of center-of-mass of alpha particles.
}
\end{figure}
In Fig.\ \ref{EHe}, calculated energy spectra of alpha particles of 
three detection angles are compared to the data \cite{SZCZUREK}.
The underestimation of the mid-energy part of the spectrum of 
$6^\circ$ and the overestimation of the cross section at $25^\circ$ 
can be again understood as to be due to the above-mentioned fact 
that the negative flow is too large in the present calculation by 
the use of the Volkov force as the effective interaction.
We have not included alpha-alpha elastic collisions as the residual 
interaction since our main interest has not been in the spectrum of 
alpha particles but in the spectra of projectile-like fragments 
such as $\Be9$.
The incorporation of alpha-alpha elastic collisions may be expected 
to have some effects in enhancing the high energy component of the 
spectrum of alpha particles.

\section{Summary}

In this paper, we have analyzed the mechanism of the reaction 
$\carbon+\carbon$ with the incident energy 28.7 MeV/nucleon with 
the antisymmetrized version of molecular dynamics (AMD) by paying 
attention mainly to its dependence on the stochastic collision 
process.
Mass distribution of produced fragments, especially with mass 
number smaller than 12, has turned out to be insensitive to the 
stochastic collision process, and is always well reproduced 
including the shell effect.
The momentum distribution of fragments is, on the other hand, very 
sensitive to the collision process.
With the inclusion of many-body nature expressed representatively 
by the nucleon-alpha collisions, the observed feature of the 
projectile fragmentation has been reproduced by calculation in this 
reaction with relatively low incident energy.
The low momentum component in the momentum distribution of 
fragments such as alpha particles and $\Be9$, which comes from 
central events, is sensitive to the value of the cross section of 
stochastic collisions and hence is expected to give precious 
information about the cross section of stochastic collisions.

Unfortunately we have not reproduced the data of momentum 
distribution perfectly since the adopted effective interaction 
results in too large attraction between the projectile and the 
target, and therefore the negative momentum flow is too large.
This means, in turn, that the close analysis of momentum 
distribution can give the information about effective interaction, 
or the momentum dependence and the density dependence of the mean 
field.
Especially the collective momentum flow of fragments such as alpha 
particles seems to be a very clear quantity which reflects the 
effective interaction,
compared to the flow of nucleons.
The analysis of the flow of fragments with other effective 
interactions, such as Gogny force, is an interesting future problem.

\acknowledgments
The computational calculation for this work has been financially 
supported by Research Center for Nuclear Physics, Osaka University.

\end{document}